\documentclass[11pt]{article}

\usepackage{floatflt}
\usepackage{multirow}
\usepackage{hyperref}
\usepackage{graphicx}
\usepackage{url}
\usepackage{color}
\usepackage{subfigure}
\usepackage{amssymb}
\usepackage{amsmath}
\usepackage{times}
\usepackage[pagewise]{lineno}

\def\b{\mathbf b}

\def\t{\mathbf t}

\def\x{\mathbf x}
\def\y{\mathbf y}
\def\z{\mathbf z}

\title{Conscious Intelligence Requires Lifelong\\
Autonomous Programming For General Purposes} 

\author
{Juyang Weng
\\
\normalsize{Department of Computer Science and Engineering}\\
\normalsize{Cognitive Science Program}\\
\normalsize{Neuroscience Program}\\
\normalsize{Michigan State University, East Lansing, MI, 48824 USA}\\
\normalsize{GENISAMA LLC, East Lansing, MI, 48824 USA}\\
\normalsize{E-mail:  weng@cse.msu.edu.}
}

\date{}

\begin{document} 




\maketitle 

\begin{abstract}
Universal Turing Machines \cite{Turing,Hopcroft06,Martin03} are well known in computer science but they are about manual programming for general purposes.  Although human children perform {\em conscious learning} (learning while being
conscious) from infancy \cite{Piaget52,Piaget54,Katz96,Elman97}, it is unknown that Universal Turing Machiness can facilitate not only our understanding of Autonomous Programming For General Purposes (APFGP) by machines, but  also enable early-age conscious learning.   This work reports a new kind of AI---conscious learning AI from a machine's ``baby'' time.   Instead of arguing what static tasks a conscious machine should be able to do during its ``adulthood'', this work suggests that APFGP is a computationally clearer and necessary criterion for us to judge whether a machine is 
capable of conscious learning so that it can autonomously acquire skills along its ``career path''.  The results here report new concepts and experimental studies for early vision, audition, natural language understanding, and emotion, with conscious learning capabilities that are absent from traditional AI systems.

\end{abstract}

To be conscious during adulthood, must a system be conscious from its ``infancy'' all the way into its
``adulthood''?  All animals \cite{Wynn92,Montague95} conduct lifelong APFGP but traditional AI systems do not \cite{LeCun15,Mnih15,Graves16,Gibney17}.
This work argues that consciousness should not be an optional crown jewelry for AI  \cite{Elman97,Koch18} but a necessity for credible AI.  Unconscious AI has resulted in machines that are brittle because they are unaware of themselves and the physical world around them.   Machines can become highly conscious by bootscraping its degree of consciousness through lifelong conscious learning.  

%
%
%

\begin{figure}
     \centering
      \includegraphics[width=0.75\textwidth]{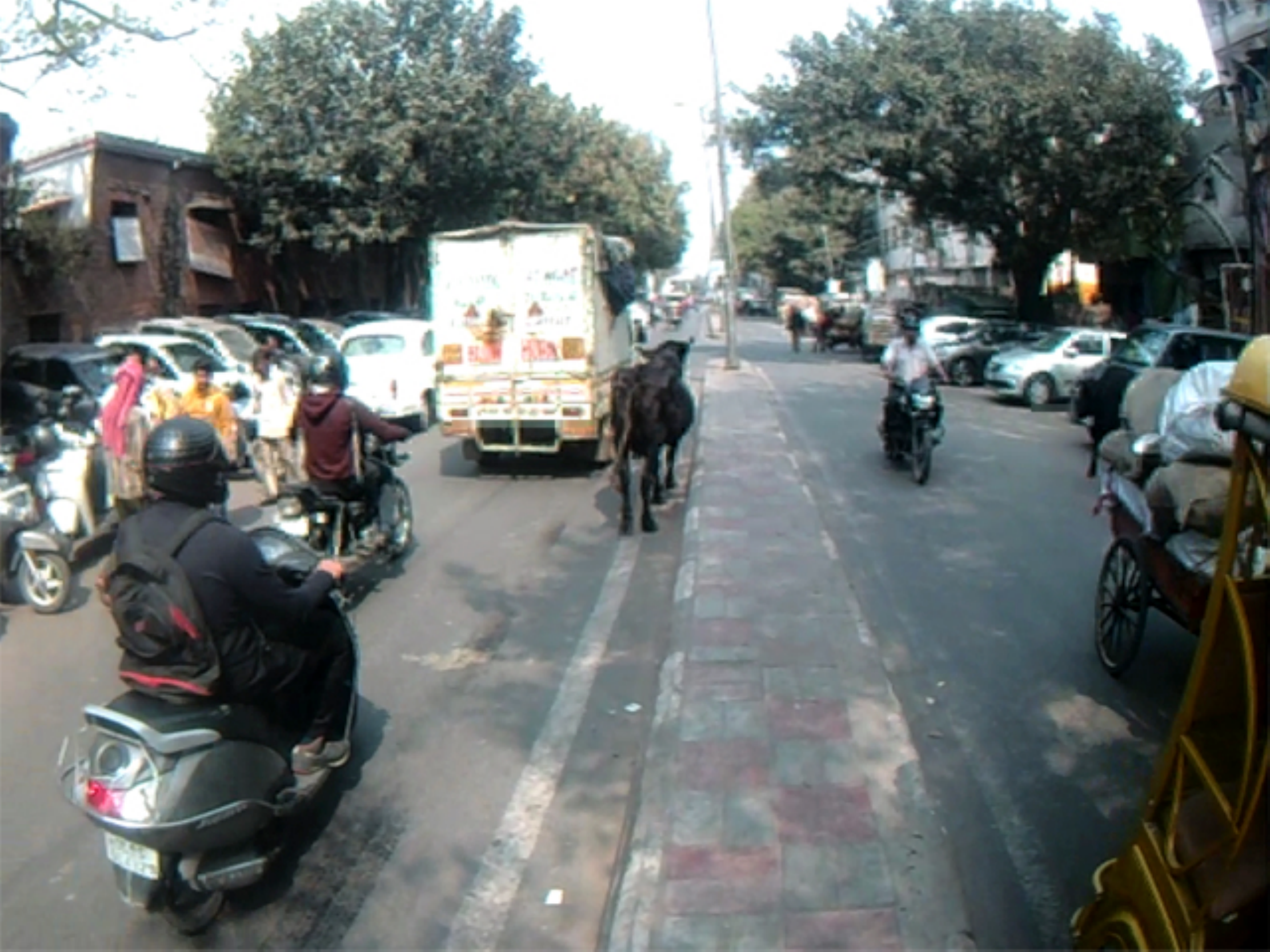}
\caption{\protect A conscious cattle and conscious humans navigate on a busy street of New Delhi, India. 
}
 \label{FG:cattle}
\end{figure}

Consciousness must apply to different contexts that a life experiences.  Furthermore, the term involves many entities, such as environment, awareness, cognition and behavior, which is learned through lifetime based on genetically pre-positioned (i.e., developed) learning capabilities \cite{Piaget52,Piaget54,Elman97}.  
For example, how does a cattle or a human in Fig.~\ref{FG:cattle} learn consciousness from infancy so that, after it has grown up, it can navigate autonomously through the hustle and bustle of streets to reach its home?  

Can an artificial machine consciously learn to do the same and more?
Weng et al. 2001  \cite{WengScience} proposed ``autonomous mental development'' as learning across lifetime that 
must be task-nonspecific.  However, the link between ``autonomous development'' and ``consciousness'' still lacks a computational minimal set.  

By ``minimal set'', we mean a minimal set of mechanisms from which a machine, natural or artificial, 
can bootstrap its degree of consciousness from ``early age'' to ``later age'' so that a degree of consciousness is already present during learning, not only after a static batch of learning.  This minimal set makes the causality of consciousness clearer.  



In order to gain understanding of what that minimal set is in terms of computation, 
let us start with a model of computation well known in computer science but not directly related to consciousness till this work.

\section{Turing Machines}
\label{SE:TM}

Turing Machines, originally proposed by Alan Turing \cite{Turing} in 1936, although not meant to explain consciousness,
can assist us to understand how consciousness arises from computations by a machine, both natural and artificial.  

\begin{figure}[tb]
     \centering
      \includegraphics[width=0.45\textwidth]{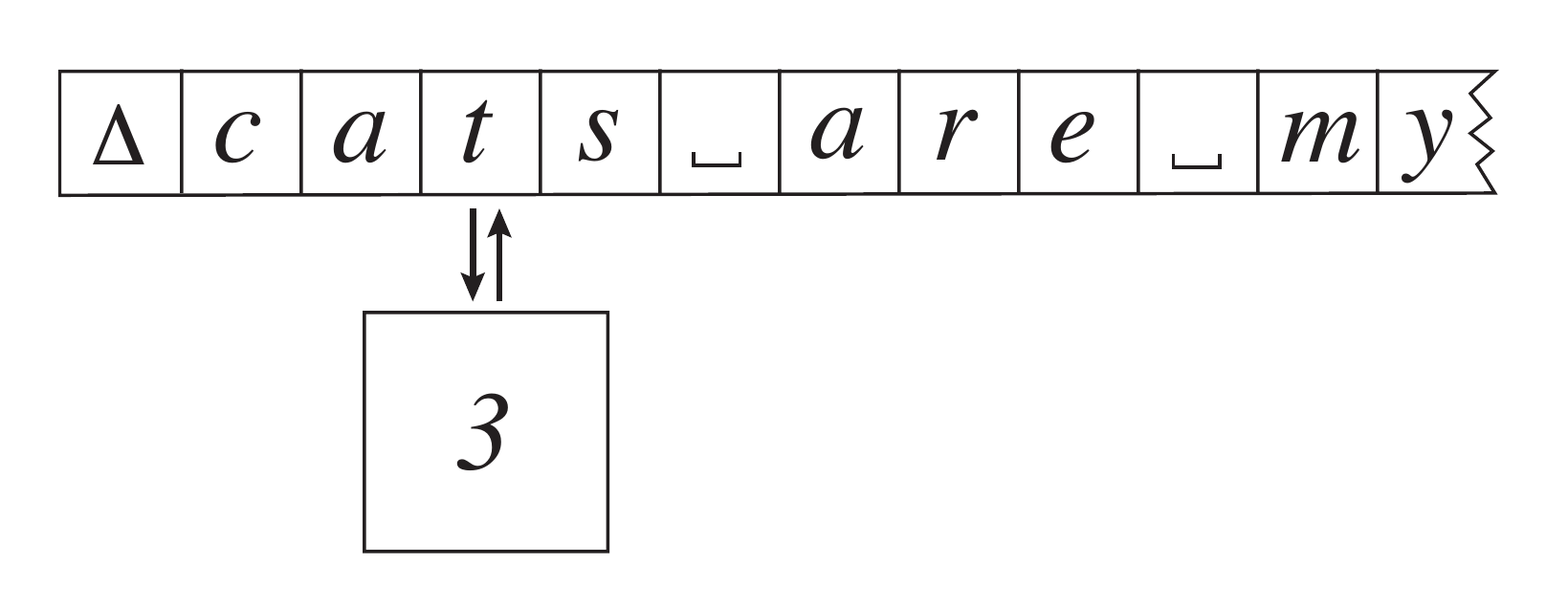}
\caption{\protect An example of Turing Machine.   Each cell of the tape bears only a symbol.  The controller has, at each integer time, a current state (e.g., 3).
}
 \label{FG:TM}
\end{figure}

A Turing Machine \cite{Hopcroft06,Martin03}, illustrated in Fig.~\ref{FG:TM}, consists of an infinite tape, a read-write head, and a controller.   The controller consists of a sequence of moves
where each move is a 5-word sentence of the following form:
 \[
 (q, \gamma ) \rightarrow (q', \gamma', d)
 \]
meaning that if the current state is $q$ and the current input that the head senses on the tape is $\gamma$, then the machine enters to next state $q'$, writes $\gamma'$ onto the tape, and its head moves in direction $d$ (left, right, or stay) but no more than one cell away.  

Intuitively speaking, let us consider each symbol in the above 5-word expression as a ``word''.
Then all such 5-word expressions are ``sentences''.   Thus, a human-handcrafted ``program'' is a sequence of such 5-word sentences the Turing Machine must follow in computation.  Although such 
sentences are not a natural language, they are more precise than a natural language.  

%
 
\section{Universal Turing Machines}
\label{SE:UTM}
How did Turing make the above machine general-purpose?   
All we need is to augment the meaning of the input on the tape:  The tape contains two parts, program $P$ and data $x$.

In his 1936 paper \cite{Turing}, Turing explained in detail how a Turing Machine can be
constructed to emulate program $P$ applied to data $x$.  
This new kind of Turing Machines is now called Universal Turing Machines \cite{Hopcroft06,Martin03}. 
We called it {\em universal} because the program $P$ on the tape is open-ended, supplied by any users for any purposes. 
However, Universal Turing Machines are still not conscious.  

To see the link between Universal Turing Machines and consciousness, we must break a series of restrictions in Turing Machines, as explained in the next section.   

\section{Eight Conditions for Consciousness}
\label{SE:GENISAMA}
The eight conditions below were not well known to be necessary for consciousness.
However, the APFGP capability in the title requires all of them.  That said, they are still insufficient for giving rise to APFGP without the full Developmental Networks (DN) to be discussed in the next section. 

To facilitate memorization, let us summarize the eight conditions in eight words: Grounded, Emergent, Natural, Incremental, Skulled, Attentive, Motivated, Abstractive, giving acronym GENISAMA.   Let us explain each of them below.

{\bf Grounded:}  Grounded means sensors and effectors of a learner must be directly grounded in the physical world in which the learner lives or operates.  IBM Deep Blue, IBM Watson, and AlphaGo are not grounded.  Instead, it is humans who synthesize symbols from the physical world, and thus shield machines 
off from the rich physical game environments, including their human opponents.  

{\bf Emergent}: The signals in the sensors, effectors and all representations inside the ``skull'' of the learner  must emerge 
automatically through interactions between the learner and the physical world by way of  sensors, effectors, and a genome (aka developmental program).  A genome is meant to fit the physical world through the entire life, not only for a specific task during the life.  For example, fruit flies must do foraging, fighting and mating.  Thus, task-specific handcrafting of representation in sensors, effectors, and inside the ``skull'' is inconsistent to consciousness.  This emergence requirement rules out task-specific and handcrafted representations, such as weights duplication in convolution used by deep learning.
Likewise, an artificial genetic algorithm without lifetime learning/development does not have anything to emerge since each individual does not learn/develop in lifetime. 

{\bf Natural}:  The learner must use natural sensory and natural motor signals, instead of human hand-synthesized features from sensors or hand-synthesized class labels for effectors, because such symbols and labels are not natural without a human in the loop.  For robots, natural signals are those directly available from 
a sensor (e.g., RGB pixel values from a camera) and raw signals for an effector/actuator.  IBM Deep Blue, IBM Watson and AlphaGo all used handcrafted symbols for the board configurations and symbolic labels for game actions.  Such symbols are not natural, not directly from cameras and not directly for robot arms.

{\bf Incremental}:  Because the current action from the learner will affect the next input to the learner (e.g., the current action ``turn left'' allows it to see 
left view), learning must take place incrementally in time.   IBM Deep Blue, IBM Watson and AlphaGo have used a batch learning method: all game configurations are available as a batch for the learner to learn.   The learner is not conscious of how it has improved from early mistakes---not self-conscious. 

{\bf Skulled}:  The skull closes the brain of the learner so that any teacher's direct manipulations with the 
internal brain representations (e.g., twisting internal parameters) are not permitted.   
For example, how can the brain be aware of what a neurosurgeon did inside the skull?

{\bf Attentive}: The learner must learn how to attend to various entities in its environment --- the body and extra-body environment.  The entities include location (where to attend), type (what to attend), scale to attend (e.g., body, face, or nose), as well as abstract concepts that the learner learned in life (e.g., am I doing the right thing?).   IBM Deep Blue, IBM Watson and AlphaGo did not seem to think ``what am I doing?''. 
 
{\bf Motivated}:  The beautiful logic that a Universal Turing Machine possesses to emulate any valid program does not give rise to consciousness as we know it.   By motivation, we mean that the learner must learn motivation based on its intrinsic motives, such as
pain avoidance, pleasure seeking, uncertainty awareness, and sensitivity to novelty.  A system that 
is designed to do facial recognition does not have a motive to do things other than facial recognition.    IBM Deep Blue, IBM Watson and AlphaGo did not feel happy when they won a game.   
  
{\bf Abstractive}:  Although a shallow definition of consciousness means awareness, full awareness requires a general capability to abstract higher concepts from concrete examples.
By higher concepts here we mean those concepts that a normal individual of 
a species is expected to be able to abstract.  Consider movie ``Rain Man'':  If a kiss by a lady on the lip is sensed only as ``wet'', there is a lack of abstraction.  A baby cannot abstract love
from the first kiss, but a normal human adult is able to.  Thus, abstraction requires learning. 

With the above eight requirements, we are ready to discuss GENISAMA Universal Turing Machines 
as super machines capable of conscious learning.

\section{GENISAMA Super Universal Turing Machines}
\label{SE:GENISAMA-UTM}

Handcrafting a University Turing Machine is not hard.  What is really hard is how to enable such a machine to
grow automatically from the natural physical word so as to learn any programs and any data directly from 
its physical environment!   We will see below how. 

\begin{figure}
     \centering
      \includegraphics[width=0.85\textwidth]{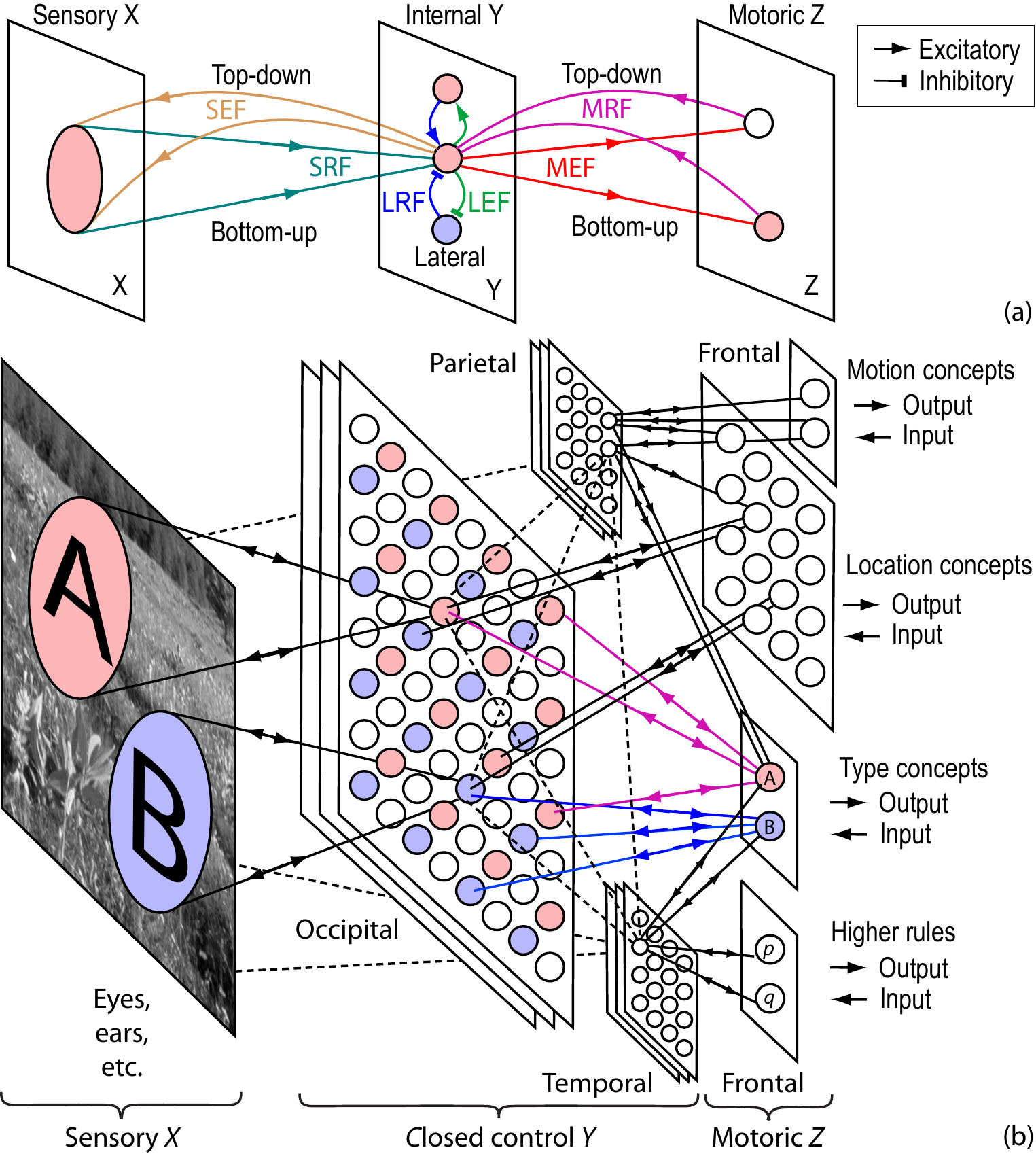}
\caption{\protect  A DN in (b) grows a brain $Y$ as a two-way bridge between the sensory bank $X$ 
and the motor bank $Z$.   All the connections are learned, updated and trimmed automatically by DN.
(a) A neuron's connections are highly recurrent which requires our time-unfolded explanation.  
}
 \label{FG:SWWnetFunctions3-2}
\end{figure}

A DN in Fig.~\ref{FG:SWWnetFunctions3-2} is capable of learning any GENISAMA Universal Turing Machine.  It grows one neuron at a time, to learn moves incrementally, one at a time.  Such a machine is capable of APFGP, which motivates us as an alternative characterization of consciousness. 

%
%


\begin{table}
\caption{Unfolding Time for APFGP in a Developmental Network}
\begin{center}
\begin{tabular}{c|cccccccccc}
\hline
Time & 0 & 1 & 2 & 3 & 4 & 5 & 6 & 7 & ... & t \\
Actable world $W$ & $W(0)$ & $W(1)$ & $W(2)$ & $W(3)$ & $W(4)$ & $W(5)$ & $W(6)$ & $W(7)$ & ... & $W(t)$  \\ 
Motor $Z$ & $Z(0)$ & $Z(1)$ & $Z(2)$ & $Z(3)$ & $Z(4)$ & $Z(5)$ & $Z(6)$ & $Z(7)$ & ... & $Z(t)$  \\ 
Skulled brain $Y$ & $Y(0)$ & $Y(1)$ & $Y(2)$ & $Y(3)$ & $Y(4)$ & $Y(5)$ & $Y(6)$ & $Y(7)$ & ... & $Y(t)$  \\ 
Sensor $X$ & $X(0)$ & $X(1)$ & $X(2)$ & $X(3)$ & $X(4)$ & $X(5)$ & $X(6)$ & $X(7)$ & ... & $X(t)$  \\ 

Sensible world $W'$ & $W'(0)$ & $W'(1)$ & $W'(2)$ & $W'(3)$ & $W'(4)$ & $W'(5)$ & $W'(6)$ & $W'(7)$ & ... & $W'(t)$  \\ 
\hline
\end{tabular}
\end{center}
\label{TB:time}
\end{table}

A brain is highly recurrent, meaning a neuron sends signals to many other neurons but other neurons also 
send signals back, directly and indirectly.  This recurrence has caused great difficulties in our understanding how the brain works.   We must unfold time
so that the time-unfolded brain is not recurrent along the time axis.   We consider five entities $W, Z, Y, X, X'$ at times $t, t=0, 1, 2, ... $, as illustrated in
Table~\ref{TB:time}.  

The first row in Table~\ref{TB:time} gives the sample times $t$.

 The second row denote the
actable world $W$, such as a hammer acting on a nail.

The third row is the motor $Z$, which has muscles to drive effectors, such as arms, legs, and mouth.

The fourth row is the skull-closed brain $Y$.  The computation inside the brain must be fully autonomous \cite{WengScience} without pre-given any tasks.

The fifth row is the sensor $X$, such as cameras, microphones, and touch sensors (e.g., skin). 

The last row is the sensible world $W'$, such as surfaces of objects that reflect light received by cameras. 

The actable world $W$ is typically not exactly the same as the sensible world $W'$, because
where sensors sense from and where effectors act on can be different.  

Next, we discuss the rules about how a DN, denoted as $N=(X, Y, Z)$, learns from world $W$ and $W'$.

Extend the tape of the Turing Machine to record images $X$ from sensors, instead of symbols $\sigma$.  
Let $X$ be the original emergent version of input, e.g., a vector that contains values of all pixels. 

Extend the output from the Turing Machine $(q', \gamma' , d)$ to be a muscle image from motor $Z$, instead of symbols.  Thus, the GENISAMA Turing Machine directly acts on the physical world. 

{\bf  Parallel computing model:}  We treat $X$ and $Z$ as external because they can be ``supervised'' by the
physical environment as well as ``self-supervised'' by the network itself.   The internal area $Y$ is 
closed (hidden)---cannot be directly supervised by external teachers. As in the above Table, we unfold the time $t$ and allow the network to have three areas $X(t)$, $Y(t)$, and $Z(t)$ that learn incrementally through time $t=0, 1, 2, ...$:
\begin{equation}
\begin{bmatrix}
Z(0)\\
Y(0)\\
X(0)
\end{bmatrix}
\rightarrow
\begin{bmatrix}
Z(1)\\
Y(1)\\
X(1)
\end{bmatrix}
\rightarrow
\begin{bmatrix}
Z(2)\\
Y(2)\\
X(2)
\end{bmatrix}
\rightarrow
...
\label{EQ:emergence}
\end{equation}
 where $\rightarrow$ means neurons on the left adaptively links to the neurons on the right.  
 
Note, all neurons in every column $t$ use only the values of the column $t-1$ to its immediate left, but use nothing from other columns.  This is true for all columns $t$, with integers $t\ge 1$.   Otherwise,  iterations are required.  Namely, by unfolding time in the above expression, the highly recurrent operations in recurrent DN become nonrecurrent in time-unfolded DN.  Thus, the DN runs in real time and do not have to slow down waiting for any iterations. 

Using Eq.~\eqref{EQ:emergence}, we outline each of the motor area $Z$, brain area $Y$, and sensory area $X$,  for $t=1, 2, ... $, from an embryo, all the way to an adult, till possible death. 

1. The motor area $Z$ starting from $Z(0)$, represents many muscles signals in a developing body.   Muscle cells in $Z(t)$ at time $t$ take inputs from the $Y(t-1)$ area and the $Z(t-1)$, acting on the environment at time $t$ using $Z(t)$ through self-supervision---trials, errors, and practices.  

2. Concurrently, the brain $Y$, starting from $Y(0)$, also dynamically develops and grows. Each neuron in 
$Y(t)$ gets multiple inputs from all three areas, $X(t-1)$, $Y(t-1)$ and $Z(t-1)$.  Competition among neurons allows only few $Y$ neurons to win.   These winner $Y$ neurons (like an expert team) at the time $t$ column represent the firing of the brain at time $t$.

3. Likewise, the sensory area $X$, starting from $X(0)$, also develops within a developing body.   What is different between the motor area $Z$ and the sensory area $X$ is that the 
former develops neurons that drive muscles (and also ``feel'' the world) but the latter develops receptors that sense the world.  

Now, we have the minimal set of mechanisms---Eq.~\eqref{EQ:emergence} along with above paragraphs 1, 2 and 3---as the {\em Computational Model of Emergence of Consciousness} for natural and artificial machines to learn consciousness (and consciously learn) throughout lifetime:
\begin{quotation}
\em As time goes by, the learner looks more and deeper aware of the world and itself by incrementally learning a lifetime program $P$ along with lifetime data  $x$ (where $P$ and $x$ are not necessarily separated) from its world, while an optimal GENISAMA Universal Turing Machine emerges as proven mathematically in \cite{WengIJIS15}.  Inside the brain this machine autonomously and recursively generates (i.e., predicts, thinks, or dreams about), at each time $t$, for the next sensory input $X$, the next brain response $Y$ and the next motor $Z$ to make a larger, more sophisticated and increasingly integrated program $P$ and to apply the program to world as $x$ for an open-ended variety of purposes.   
\end{quotation}
In the eyes of humans, this learner becomes increasingly conscious.  
Although the lifelong-learned consciousness can be extremely complex, the minimal set of computational mechanisms is relatively simple for us to understand, thanks to the Universal Turing Machines.

The remaining detail of this work is discussed in Methods.  See \cite{WengJCS2020} for a less mathematical paper about APFGP oriented towards cognitive scientists and neuroscientists.
\section{Methods}

\subsection{Developmental Networks for Learning Consciousness}  
\label{SE:DN}

A Developmental Network  is meant for consciousness because it is a holistic model for a biological brain, also fully implementable on an artificial machine. 

The following section presents Developmental Network 1 (DN-1).  Developmental Network 2 (DN-2) 
is
different from DN-1 primarily in the following sense.  In DN-1, each of multiple $Y$ areas has a static set of neurons so that the competition within each area is based on a top-$k$ principle.  Namely, inhibition
among neurons within each area is implicitly modeled by top-$k$ competition.  

In the DN-2, however, there is no static assignment of neurons to any regions, so that regions in
the DN-2 automatically emerge, along their scales, cascade, and nesting.   A major advantage of DN-2
is that a human programmer is not in the loop of deciding the distribution of X-Y-Z mechanisms, relieving
human from this intractable task of handcrafting consciousness.  A major disadvantage of DN-2 is that its computational
explanation is too sophisticated to be included in this paper.   Let us leave DN-2 out from this paper and concentrate on DN-1 below. 

The hidden $Y$ area corresponds to the entire ``brain''.   In the following, we assume the brain has a single
area $Y$ but it will enable many subareas to emerge.

The brain takes input from vector $(\z, \x)$, not just sensory $\x$ but also motor $\z$, to produce an internal response vector $\y$ which represents the best match of $(\z, \x)$ with one of many 
internally stored patterns of $(\z, \x)$:
 
The winner-take-all learning rule, which is highly nonlinear and simulates parallel lateral inhibition in the internal (hidden) area $Y$ is sufficient to prove in \cite{WengIJIS15} that a DN that 
has sufficient hidden neurons learns any Turing Machine perfectly, immediately, and error-free.   

The $n$ neurons in $Y$ give a response vector $\y= (y_1, y_2, ... y_n)$ of $n$ neurons in which only
the best-matched neuron fires at value 1 and all other neurons do not fire giving value 0: 
\begin{equation}
y_j = \begin{cases} 1 & \mbox{if $j=\underset{1\le i \le n}{\operatorname{argmax}} 
\{ f( \t_i, \z, \b_i, \x )\}$}\\
0 & \mbox{otherwise}
\end{cases}
\;\;\;\;j=1, 2, ... n,
\label{EQ:Ycompute}
\end{equation}
where $f$ is a function that measures the similarity between the top-down 
weight vector $ \t_i $ and the top-down input vector $\z$ \cite{Saalmann07} as well as the similarity between the bottom-up weight vector $\b_i $ and the bottom-up input vector $\x$.  The value of similarity is the inner product of their length-normalized versions \cite{WengIJIS15}.  Corresponding to FA, both the top-down weight and the bottom-up weight must match well for $f$ to give a high value as inner product.  

The response vector $\y$ the hidden $Y$ area of DN is then used by $Z$ and $X$ areas to predict the next $\z$ and $\x$ respectively in discrete time $t = 1, 2, ,3, ... $:
\begin{equation}
\left[
\begin{array}{c}
  \z(t-1)  \\
  \x(t-1)   
\end{array}
\right]
\rightarrow
\y(t)
\rightarrow
\left[
\begin{array}{c}
  \z(t+1) \\
  \x(t+1)
\end{array}
\right]
\label{EQ:DNtransition}
\end{equation}
where $\rightarrow$ denotes the update on the left side using the left side 
as input.  The first $\rightarrow$ above is highly nonlinear because of 
the top-1 competition so that only one $Y$ neuron fires (i.e., exactly one component in binary $\y$ is 1).
The second $\rightarrow$ consists of simply links from the single firing $Y$ neurons to 
all firing neurons on the right side.   

Like the transition function of a Turing Machine, each prediction of $\z(t+1)$ in  Eq.~\eqref{EQ:DNtransition} is called a {\em transition}, but now in real-valued vectors without any symbols.  The same $\y(t)$ can also be used to predict the binary (or real-valued) $\x(t+1)\in X$ in Eq.~\eqref{EQ:DNtransition}.   The quality of prediction of $(\z(t+1) , \x(t+1))$ 
depends on how state $Z$ abstracts the external world sensed by $X$.  The more mature the DN
is in its ``lifetime''' learning, the better its predictions. 

The expression in Eq.~\eqref{EQ:DNtransition}, is
extremely rich as illustrated in Fig.~\ref{FG:SWWnetFunctions3-2}: Self-wiring within a Developmental Network (DN) as the control of GENISAMA TM, based on statistics of activities through ``lifetime'', without any central controller,  Master Map, handcrafted features, and convolution.  

The above vector formalization is simple but very powerful in practice. 
The pattern in $Z$ can represent the
binary pattern of any abstract concept --- context, state, muscles, action, intent, object type, object 
group, object relation.  However, as far as DN is concerned, they mean the same--- a firing pattern of the $Z$ area! 

Namely, unified numerical processing-and-prediction in DN amounts to any abstract concepts above. 
In symbolic representations, it is a human to handcraft every abstract concept as a symbol; but DN does not  have a human in the ``skull''. it simply learns, processes, and generates vectors.  In the eyes of a human 
outside the ``skull'', the DN gets smarter and smarter.  

Eq.~\eqref{EQ:DNtransition}(a) shows each feature neuron has 
six fields in general: Sensory Receptive Field (SRF), Sensory Effective Field (SEF), Motor Receptive Field (MRF), Motoric Effective Field (MEF), and Lateral Receptive Field (LRF) and Lateral Effective Field (LEF).
Eq.~\eqref{EQ:DNtransition}(b) shows the resulting self-wired architecture of DN with Occipital, Temporal, Parietal, and Frontal lobes.  Regulated by a general-purpose Developmental Program (DP), the DN self-wires by ``living'' in the physical world.  The $X$ and $Z$ areas are supervised by body and the physical world which includes teachers.  
 
 Through the synaptic maintenance, some $Y$ neurons 
gradually lose their early connections (dashed lines) with $X$ ($Z$) areas and become ``later'' (early) $Y$ areas.  In the (later) Parietal and Temporal lobes, some neurons further gradually lost their connections with the (early) Occipital area and become rule-like neurons.  These self-wired connections give rise to a complex dynamic network, with shallow and deep connections instead of a deep cascade of areas.  
Object location and motion are non-declarative concepts and object type and language sequence
are declarative concepts.  Concepts and rules are abstract with the desired specificities and invariances. DN does not have any static Brodmann areas.

\subsection{Optimal Properties Proved for DN}
\label{SE:properties}

If a DN can learn quickly like other normal animals, we may have to call it retarded with only a limited
consciousness compared to other animals of the same age.   We do not want a DN to get stuck into a local minimum, as many nonlinear artificial systems have suffered.  

Fortunately, every DN is optimal in the sense of maximum likelihood, proven mathematically in \cite{WengIJIS15}.  Put intuitively, all DNs are optimal, given the same learning environment, the same learning experience, and the same number of neurons in the ``brain''.  There might be many possible network solutions some of which got stuck into local minima in their search for a good network.  However, each DN is the most likely one, without getting stuck into local minima. 
This is because although a DN starts with random weights, all random weights result in the same network.  

However, this does not mean that the learning environment is the best possible one or the number of neurons is best
possible one for many lifetime tasks.   Search for a better educational environment will be human challenge for their future children, both natural and artificial kinds. 

\subsection{Experiments}
\label{SE:exp}

This seems the first time where general-purpose vision, general-purpose audition, and general-purpose natural language, as the three well-known bottleneck areas in AI has been learned by a single type of network integrated with motivational learning.  Other systems include \cite{Gallistel99,Jenkins08,Tenenbaum11,Eliasmith12,Jordan15,Lake16,Moravcik17,Holm19}.

We conducted experiments in which a learning system acts as an emergent Turing Machine that learns one
of three well-recognized bottleneck problems in AI, vision, audition and natural language acquisition.  
Hopefully, when such systems are mature enough after ``living'' and ``learning'' in the real physical world, they look as though they have a certain degree of animal-like consciousness in the eyes of humans. 

{\bf Vision from a ``lifelong'' retinal sequence:}   How does a DN become visually conscious demonstrated by its motor behaviors?   Let it learn by artificially ``living'' in the real world!   

\begin{figure}[tb]
     \centering
      \includegraphics[width=0.98\textwidth]{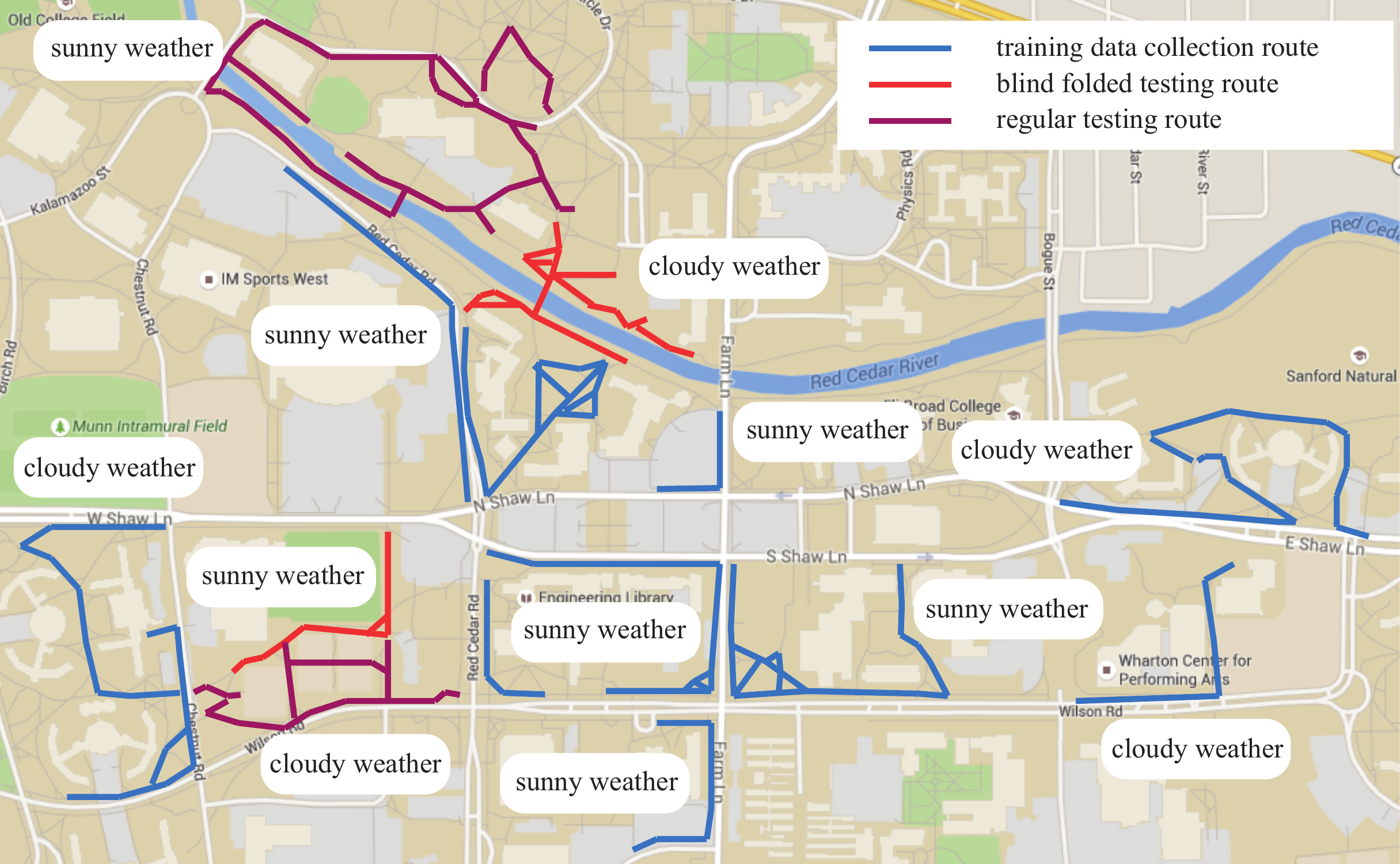}
\caption{\protect\label{FG:navigation_training_route}  
Training, regular testing, and blind-folded testing sessions conducted on campus of Michigan State University (MSU), under different times of day and different natural lighting conditions (e.g., there are extensive shadows in images).  Disjoint testing sessions were conducted along paths that the machine has not learned.  This is the first time for visual awareness to be learned by GENISAMA Turing Machines.}
\end{figure}

Fig.~\ref{FG:navigation_training_route} provides an overview of the extensiveness of the training, regular 
training, and blindfolded testing sessions.
The inputs to the DN were from the same mobile phone
that performs computation.  They include the current image from the monocular camera, the current desirable direction from the Google Map API and the Google Directions API.  If the teacher imposes
the state in $Z$, this is treated as the supervised state.  Otherwise, the
DN outputs its predicted state from $Z$.  The DN learned to attend critical visual information in the current image (e.g., scene type, road features, landmarks, and obstacles) depending on the context of desired direction and the context state.  Each state from DN includes heading direction or stop, the location of the attention, and the type of object to be detected (which detects a landmark), and the scale of attention (global or local), all represented as binary patterns.  None is a symbol.

For further detail of learning vision-guided navigation and planning for navigation see  \cite{WengIJCNN2020}.

Below, we will see that an auditory consciousness also uses the same DN, but using different ``innate'' parameters. 

{\bf Audition from a ``lifelong'' cochlear sequence:}   How does a DN become auditory conscious demonstrated by its motor behaviors?   Let it learn by artificially ``living'' in the real world!  

For the audition modality, each input image to $X$ is the pattern that simulates the output from an array of hair cells in the cochlea.  We model the cochlea in the following way.   The cells in the base of the cochlea correspond to filters with a high pass band.  The cells in the top correspond to filters with a low pass band.   At the same height, cells have different phase shifts.   Potentially, such a cochlear model could deal with music and other natural sound, more general than the popular Mel Frequency Cepstral Coefficients (MFCCs) that are mainly for human speech processing.  
The performance will be reported elsewhere due to the limited space.

\begin{figure}[tb]
\centering
 \includegraphics[width=0.78\textwidth]{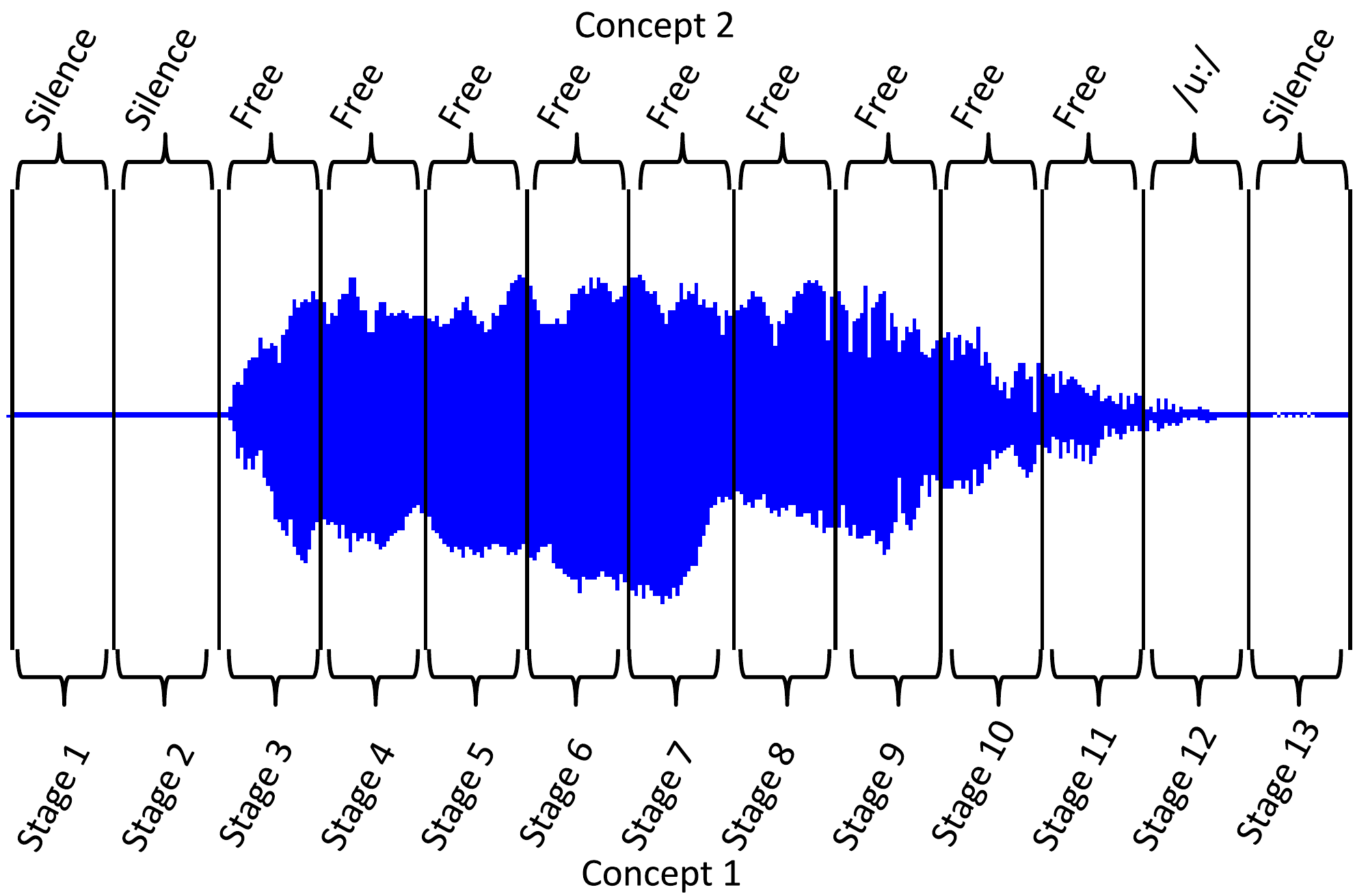}
\caption{The sequences of concept 1 (dense, bottom) and concept 2 (sparse, top) for phoneme /u:/.  The latest DNs do not need human to provide any labels.  Instead, they self-supervise themselves. 
}
\label{FG:audio-u}
\end{figure}

Take the phoneme /u:/ as an example shown in 
Fig.~\ref{FG:audio-u}. The state of concept 2 keeps as silence when inputs are silence frames. It becomes a ``free'' state when phoneme frames are coming in, and changes to /u:/ state when first silence frame shows up at the end.  At the same time, the states of concept 1 count temporally dense stages. 

For more details of auditory learning using Developmental Networks, see  \cite{WengIJCNN2020}.

One may ask, what about higher consciousness such as natural language understanding?

{\bf Natural languages from a ``lifelong'' word sequence:}   How does a DN become language conscious demonstrated by its motor behaviors?   Let it learn by artificially ``living'' in the real world!   Here, we assume grounded words are
emergent patterns, not symbols. 

As far as we know, this seems to be the first work that deals with language acquisition in a bilingual environment, largely because the DN learns directly from emergent patterns,
both in word input and in action input (supervision), instead of static symbols.   

The input to $X$ is a 12-bit binary pattern, each represents a word, which potentially can represent 
$2^{12}$ words using binary patterns.    The system was taught 1,862 English and French sentences from
\cite{Scriven11}, using $2,338$ unique words (case sensitive).  As an example of the sentences:  English: ``Christine used to wait for me every evening at the exit.'' French: ``Christine m'attendait tours les soirs \`{a} la sortie.''

\begin{figure}[tb]
\centering
 \includegraphics[width=0.78\textwidth]{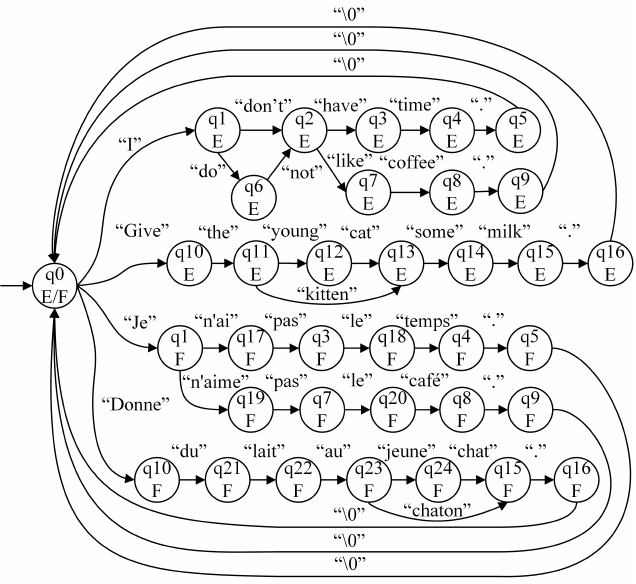}
\caption{The finite automaton for the English and French versions of some sentences.   The DN learned a much larger finite automaton.  Cross-language meanings of partial- and full-sentences are represented by the same state of meaning context $q_i$, $i=0, 1, 2,  ... , 24$.  See, e.g., $q_1$, $q_3$, $q_4$, and $q_5$.  But the language specific context is represented by another concept: language type.   The last letter is the return character that indicates the end of a sentence.}
\label{FG:bilingualFA}
\end{figure}

The $Z$ area was taught two concepts:  language type (English, French, and language neutral, e.g., a number or name)
represented by 3 neurons (top-1 firing), and the language-independent meanings as meaning states, as
shown in Fig.~\ref{FG:bilingualFA}.
The latter is represented by 18 neurons 
(18-bit binary pattern), always top 5 neurons firing, 
capable of representing $C(18,5)=8,568$ possible combinations as states, but only $6,638$ actual meanings were recorded.   Therefore, the $Z$ area has
$3+18=21$ neurons, potentially capable of representing a huge number $2^{21}$ binary patterns if
all possible binary patterns are allowed.   

However, the 
DN actually observed only $8,333$ $Z$ patterns (both concepts combined) from the training experience, and
$10,202$ distinct $(Z,X)$ patterns---FA transitions.   Consider a 
traditional symbolic FA using a symbolic transition table, which has 
$6,638\times 3 = 19,914$ rows and $2,338$ columns.  This amounts to $19,914\times 2,338=46,558,932$ table entries.  

But only  $10,202/46,558,932 \approx 0.022\%$ of the entries were detected by the hidden neurons,
representing that only $0.02\%$ of the FA transition table was observed and accommodated by the DN.
Namely, the DN has a potential to deal with 
$n$-tuples of words with a very large $n$ but bounded by DN size, because most un-observed $n$-tuples are never represented.  The FA transition table is extremely large, but never generated.  

Without adding noise to the input $X$, the recognition error is zero, provided that there is a sufficient number of $Y$ neurons.   We added Gaussian noise into the bits of $X$.   Let $\alpha$ represent the 
relative power of the signal in the noisy signal.  When $\alpha$ is 60\%, the state recognition rate
of DN is around 98\%.  When $\alpha$ is 90\%, the DN has reached 0\% error rate, again thanks to
the power of DN internal interpolation that converts a huge discrete (symbolic) problem into a 
considerably smaller continuous (numeric) problem.  

See \cite{WengIJCNN2020}
for more detail.  

{\bf Emotional learning using the same network:} One may wonder, does this type of consciousness enable emotion?  The DN model considers emotion to belong to a wider category known in neuroscience as motivation. 

Motivation is very rich \cite{Dolan02}.  It has two major aspects (a) and (b) in the current DN model.   All reinforcement-learning methods other than the DN, as far as we know, are for symbolic methods (e.g., Q-learning \cite{Sutton98,Mnih15}) and are in aspect (a) below exclusively.   The DN uses concepts (e.g., important events) instead of the rigid time-discount in Q-learning to avoid the failure of far goals.

(a) Pain avoidance and pleasure seeking to speed up learning important events.  Signals from pain  (aversive) sensors release a special kind of neural transmitters (e.g., serotonin \cite{Daw02})  that diffuse into all neurons that suppress $Z$ firing neurons but speed up the learning rates of the firing $Y$ neurons.    Signals from sweet (appetitive) sensors release a special kind of neural transmitters (e.g., dopamine \cite{Kakade02})  that diffuse into all neurons that excite $Z$ firing neurons but also speed up the learning rates of the firing $Y$ neurons.   Higher pains (e.g., 
loss of loved ones and jealousy) and higher pleasure (e.g., praises and respects) develop at later ages from lower pains and pleasures, respectively. 

(b) Synaptic maintenance---the growing and trimming of the spines of synapses--- 
segments object/event and motivates curiosity.   Each synapse incrementally estimates the average error $\beta$ between the pre-synaptic signal and
the synaptic conductance (weight), represented by a kind of neural transmitter (e.g.,  acetylcholine \cite{Yu05}). 
Each neuron estimates the average deviation $\bar\beta$ as the average across
all its synapses.   
The ratio $\beta/\bar\beta$ is the novelty represented by a kind of neural transmitters (e.g., norepinephrine, \cite{Yu05}) at each synapse.  The synaptogenic factor $f(\beta, \bar\beta)$ at each synaptic spine
and full synapse enables the spine to grow if the ratio is low (1.0 as default) and
to shrink if the ratio is high (1.5 as default).   See Fig.~\ref{FG:SWWnetFunctions3-2}(b) for how a neuron can cut off their direct connections with $Z$ to become 
early areas in the occipital lobe or their direct connections with the $X$ areas to become latter areas inside
the parietal and temporal lobes.   However, we cannot guarantee that such ``cut off'' are 100\% based on the statistics-based wiring theory here. 

See \cite{WengDNreinforcement13,Wang11,Guo15} 
for more details about motivational learning in DN.

\subsection{Conclusions}
\label{SE:conclusions}
APFGP inside a network has a minimal set of computational mechanisms for conscious systems, natural and artificial.   The new APFGP characterization is clearer than existing other characterizations for the notoriously vague term ``consciousness'' as we discussed in the first section.  Hopefully, APFGP will give rise to richer animal-like artificial consciousness so that conscious AI receives a long-overdue credibility for AI.  
APFGP might also be useful as a computational model for unifying natural consciousness and artificial consciousness, due to its holistic nature backed by
the new capability---APFGP of GENISAMA Universal Turing Machines.  Much exciting practical work on learning consciousness remains to be done in the future, including creating conscious AI and verifying 
APFGP on natural conscious systems.   This is a constructive proof that natural consciousness is computational and does not need quantum mechanics, contrary to the quantum hypothesis by Roger Penpose \cite{PenroseSM}.  Hopefully, artificial consciousness would 
follow this methodology to approach human levels.




\section*{Addendum}
\begin{description}
\item[Acknowledgement:] The author likes to thank Zejia Sheng, Xiang Wu and Juan Castro-Garcia for conducting experiments 
which will be further reported in \cite{WengIJCNN2020}.
\item[Competing Interests] The author declares that he has no
competing financial interests.
\item[Correspondence] Correspondence and requests for materials
should be addressed to Juyang Weng (email: juyang.weng@cse.msu.edu).
\end{description}



\end{document}